\documentclass[aps,prl,twocolumn,showpacs,superscriptaddress]{revtex4}
\begin{document}
\title{Density of states of the interacting two-dimensional electron gas}
\author{E. Kogan}
\email{kogan@quantum.ph.biu.ac.il}
\affiliation{Jack and Pearl Resnick Institute, Physics
Department, Bar Ilan University, Ramat Gan  52900, Israel}
\affiliation{National Center for Theoretical Sciences and Electrophysics Department,
National Chiao Tung University, Hsinchu 30050, Taiwan}
\author{B. Rosenstein}
\email{baruchro@hotmail.com}
\affiliation{National Center for Theoretical Sciences and Electrophysics Department,
National Chiao Tung University, Hsinchu 30050, Taiwan}
\affiliation{Jack and Pearl Resnick Institute, Physics
Department, Bar Ilan University, Ramat Gan  52900, Israel}
\date{\today}

\begin{abstract}
We study the influence of electron-electron interactions 
on the density of states (DOS) of clean
2D electron gas. We confirm the  linear cusp in the DOS
around the Fermi level, which was obtained previously. The  cusp crosses over 
to a pure logarithmic dependence further away from the Fermi surface. 

\end{abstract}
\maketitle

It was established more than 20 years ago by Altshuler and Aronov \cite{alt1} and
Altshuler, Aronov,  and  Lee \cite{alt2} that  in low-dimensional diffusive systems the
electron-electron interaction leads to suppression of the single-particle
density of states  (DOS) at the Fermi level. In two dimensions their theory predicted
logarithmic cusp at the 
Fermi level due to the diffuson pole divergences in the vertex renormalizaion. 
The theory was later
extended  by Rudin, Aleiner and Glazman \cite{rudin} to larger values of the energy 
measured from the Fermi surface. 

Recently it was shown 
by Khveshchenko and Reizer \cite{khv} and  by Mishchenko and Andreev \cite{mish} 
using the diagrammatic approach with the RPA dynamical susceptibility,
that large  electron-electron interaction induced correction to the
DOS  exists even in the absence of
disorder.  Both groups obtained a linear cusp  
at the Fermi level 
$\delta\nu (\epsilon)/\nu _{0}\sim\left\vert \epsilon \right\vert /E_{F}$ 
(with the slope differing by a factor of 2),
independent of the strength of the
electron - electron interaction $r_{s}$. 
More recently Rollbuhler and Grabert  \cite{roll}  
using a path integral technique (applied for relatively small couplings $r_{s}<1$) 
found numerically that the slope does depend on the coupling and flattens away from the Fermi 
surface.

Due to apparent  discrepancy between the two 
approaches we reconsider the problem  in the framework of the diagrammatic method.
Our aim is to spell explicitly and to clarify all the approximations which are made 
to obtain the transparent analytical formulas for the  DOS correction.
We also obtain the results which are partially different from those of  
Refs. \onlinecite{khv,mish}.

The Hamiltonian of the 2D electron gas is
\begin{equation}
\label{hamiltonian}
H=\sum_{p\sigma }\varepsilon _{p}a_{p\sigma }^{\dagger }a_{p\sigma }
+\frac{1}{2}\sum_{pp^{\prime }\sigma \sigma ^{\prime }q}a_{p+q\sigma }^{\dagger}a_{p^{\prime }
-q\sigma ^{\prime }}^{\dagger }V(q)a_{p^{\prime }\sigma^{\prime }}a_{p}\sigma ,
\end{equation}
where
\begin{equation}
\varepsilon _{p}=\frac{p^{2}}{2m};\;V(q)=\frac{2\pi e^{2}}{q}.
\end{equation}
The Green's function $G(p,E)$  is given by the equation
\begin{equation}
G^{-1}(p,E)=G_{0}^{-1}(p,E)-\Sigma (p,E),
\end{equation}
where $G_{0}(p,E)$ is Green's function in the 
absence of electron-electron interaction
\begin{equation}
G_{0}^{-1}(p,E)=E -\frac{p^{2}}{2m};
\end{equation}
the self-energy $\Sigma (p,E)$  in this paper will be
calculated in the Random Phase approximation (RPA) at $T=0$:
\begin{equation}
\Sigma (p,E)=\int \frac{d^{2}q}{^{\left( 2\pi \right) ^{2}}}\int \frac{%
d\omega }{^{2\pi }}G_{0}(p+q,E+\omega )\frac{V(q)}{\varepsilon (q,\omega)}.
\end{equation}
The DOS is
\begin{equation}
\nu (E)=-\frac{2}{\pi }{\rm Im}\int 
\frac{d^{2}p}{^{\left( 2\pi\right) ^{2}}}G(p,E+i0).
\end{equation}
Expanding Green's function with respect to the self energy and reversing the
order of integrations, one obtains the electron-electron interaction 
induced correction to the DOS  
\cite{alt1}
\begin{equation}
\delta \nu (E)=-\frac{2}{\pi }{\rm Im}\int 
\frac{d^{2}p}{^{\left( 2\pi\right) ^{2}}}\delta G(p,E+i0)\equiv
{\rm Im} X(E+i0).
\end{equation}
We will work in Matsubara formalism, calculate $X$ for imaginary frequency
\begin{eqnarray}
\label{deltax}
X(i\Omega)=-\frac{2}{\pi }\int \frac{d^{2}q}{^{\left( 2\pi \right) ^{2}}}
\int_{-\infty }^{\infty }\frac{d\omega }{2\pi }\frac{V(q)}
{\varepsilon(q,i\omega )}\nonumber\\
\int \frac{d^{2}p}{^{\left( 2\pi \right)^{2}}}G_{0}^{2}
(p,i\Omega)G_{0}(p+q,i(\Omega+\omega )),
\end{eqnarray}
and
at the end will make an analytical continuation.

Using the fact that $G_{0}^{2}(p,i\Omega)=i\partial G_{0}(p,i\Omega)/\partial
\Omega$, we can
write the integral over $p$ of the three Green's functions  as 
\begin{eqnarray}
\int \frac{d^{2}p}{^{\left( 2\pi \right)^{2}}}G_{0}^{2}(p,i\Omega)
G_{0}(p+q,i(\Omega+\omega ))\nonumber\\
=-i\left( \frac{\partial }{\partial \Omega}
-\frac{\partial }{\partial \omega }\right) \Pi (q,i\omega ,i\Omega),
\end{eqnarray}
where the polarization function 
$\Pi (\Omega ,q,\omega )$ is given by the equation
\begin{eqnarray}
\label{pol}
\Pi (q,i\omega ,i\Omega )=\int \frac{d^{2}p}{^{\left( 2\pi\right)^{2}}}
G_{0}(p,i\Omega)
G_{0}(p+q,i(\Omega+\omega )).
\end{eqnarray}
The last equation
in polar coordinates takes the form
\begin{eqnarray}
\Pi (q,i\omega ,i\Omega )=\frac{1}{\left( 2\pi \right) ^{2}}\int_{0}^{2\pi
}d\theta \int_{0}^{\infty }pdp\nonumber\\
\frac{1}{i\Omega +\mu -\frac{p^{2}}{2m}}\;\;\;
\frac{1}{i(\Omega +\omega )+\mu -\frac{p^{2}}{2m}-\frac{pq\cos (\theta )}{m}
-\frac{q^{2}}{2m}}.
\end{eqnarray}
Since the main contributions come from fermionic momenta close to Fermi
momentum, one can make the first approximation by replacing $p$ in the term 
$pq\cos(\theta)/m$ by 
$\sqrt{2m\mu }=mv_{F}$ and ignoring the term $q^{2}/2m$.  
After that the integration over $p$
can be easily performed:
\begin{eqnarray}
\label{ppp}
\Pi (q,i\omega ,i\Omega )=\frac{m}{(2\pi )^{2}}\int_{0}^{2\pi }
\frac{d\theta }{i\omega -v_{F}q\cos (\theta )}L(\omega,\Omega,q),
\end{eqnarray}
where 
\begin{equation}
L(\omega,\Omega,q)=\log \frac{(\Omega +\omega )-i\mu
+iv_{F}q\cos (\theta )}{\Omega -i\mu }.
\end{equation}
Explicitly, presenting the logarithm as
\begin{eqnarray}
L(\omega,\Omega,q)=\frac{1}{2}\log \frac{(\Omega +\omega )^{2}
+(\mu-v_{F}q\cos(\theta ))^{2}}{\Omega ^{2}+\mu ^{2}}\nonumber\\
+i\left[ \tan ^{-1}\frac{\mu }{\Omega }
-\tan ^{-1}\frac{\mu-v_{F}q\cos (\theta ) }{\Omega +\omega }\right] ,
\end{eqnarray}
and taking into account that $\Omega,\omega,v_{F}q \ll \mu$, we
make the second approximation:
\begin{eqnarray}
L=\pi i\left[ \Theta (-\Omega )\Theta (\Omega
+\omega )-\Theta (\Omega )\Theta (-\Omega -\omega )\right].
\end{eqnarray}
After that, the 
integral over $\theta$ in Eq. (\ref{ppp}) can be easily calculated, and
the  polarization function takes a form
\begin{eqnarray}
\label{polf}
\Pi (q,i\omega ,i\Omega )=\frac{m}{2}\frac{\text{sign}(\omega )}{\sqrt{\omega ^{2}+v_{F}^{2}q^{2}}}\nonumber\\
\left[ \Theta (-\Omega )\Theta (\Omega
+\omega )-\Theta (\Omega )\Theta (-\Omega -\omega )\right].
\end{eqnarray}
Taking appropriate derivatives one obtains 
\begin{eqnarray}
\label{3g}
&&\int \frac{d^{2}p}{^{\left( 2\pi \right) ^{2}}}
G_{0}^{2}(p,i\Omega)G_{0}(p+q,i(\Omega+\omega )) =\frac{-m|\omega |i}{2[\omega ^{2}+v_{F}^{2}q^{2}]^{3/2}}
\nonumber\\
&&\left[ \Theta(-\Omega )\Theta (\Omega +\omega )-\Theta (\Omega )\Theta (-\Omega -\omega )\right] .
\end{eqnarray}

The dielectric constant $\varepsilon (q,\omega )$ is related to the
polarization operator
\begin{eqnarray}
\label{p}
P(q,i\omega )=-2\int \frac{d\Omega }{2\pi }\Pi (q,i\omega ,i\Omega )
\end{eqnarray}
by the equation
\begin{equation}
\varepsilon (q,i\omega )=1-V(q)P(q,i\omega ).
\end{equation}
It is  easier, however, not to use
the approximate Eq. (\ref{polf}) for the polarization function,
but to insert in Eq. (\ref{p})  exact Eq. (\ref{pol}) and 
(as it is traditionally done)  
integrate 
over  $\Omega$ first, to obtain
\begin{equation}
P(q,i\omega )=2\sum_{p}\frac{n_{p}-n_{p+q}}{\varepsilon _{p+q}-\varepsilon_{p}-i\omega },
\end{equation}
where $n_{p}$ is the Fermi distribution function.  Obvious
algebra gives
\begin{eqnarray}
&&P(q,i\omega ) =\frac{1}{\pi ^{2}}{\rm Re}\int_{p<p_{F}}pdp\int d\theta 
\frac{1}{\frac{pq\cos \theta }{m}+\frac{q^{2}}{2m}-i\omega }\nonumber\\
&&=\frac{2}{\pi }{\rm Re}\int_{p<p_{F}}pdp\frac{1}{\sqrt{\left( \frac{q^{2}}{2m}-i\omega
\right) ^{2}-\frac{p^{2}q^{2}}{m^{2}}}}
\end{eqnarray}
Integrating over  $p$, 
\begin{eqnarray}
P(q,i\omega ) =\frac{2m^{2}}{\pi q^{2}}\nonumber\\
{\rm Re}\left[ \frac{q^{2}}{2m}-i\omega -\sqrt{%
\left( \frac{q^{2}}{2m}-i\omega \right) ^{2}-v_{F}^{2}q^{2}}\right],
\end{eqnarray}
and expanding the radical in small $q$, one obtains
\begin{equation}
\label{diel}
\varepsilon (q,i\omega )=1+\frac{2e^{2}m}{q}\left[ 1-\frac{|\omega|}
{\sqrt{\omega ^{2}+v_{F}^{2}q^{2}}}\right].
\end{equation}

Now we are ready to return to the calculation of the DOS. Substituting Eqs.(\ref{3g})
and ({\ref{diel}) into Eq.(\ref{deltax}), considering for definiteness $\Omega>0$, 
and subtracting
an "inessential" constant, one has
\begin{eqnarray}
\label{x}
X(i\Omega)=\frac{e^{2}mi}{2\pi ^{2}}\int_{0}^{\Omega}
\omega d\omega\int_{0}^{\infty}\frac{qdq }{[\omega ^{2}+v_{F}^{2}q^{2}]^{3/2}} \nonumber\\
\frac{1}{q+2e^{2}m\left[1-\frac{\omega}{\sqrt{\omega^{2}+v_{F}^{2}q^{2}}}\right]}.
\end{eqnarray}

Apart
from the fact that we are considering imaginary energy, 
Eq. (\ref{x}) coincides with those obtained in 
Refs. \onlinecite{khv,mish}. (Comparing with the Eq. (2) from the 
latter reference, 
we have an additional factor $1/2$.) We notice, however,
that the double integral can be calculated in a more rigorous way than it was
done there.
After we introduce the dimensionless variable 
$\bar{q}=v_{F}q/\omega$, and change the order of integrations, 
the integral takes a form
\begin{eqnarray}
X(i\Omega)=\frac{e^{2}mi}{2\pi ^{2}}
\int_{0}^{\infty }\frac{\bar{q}d\bar{q}}{[1+\bar{q}^{2}]^{3/2}}\nonumber\\
\times\int_{0}^{\Omega}\frac{d\omega}{\frac{\bar{q}}{v_F}\omega +2e^{2}m
\left[1-\frac{1}{\sqrt{1+\bar{q}^{2}}}\right]}.
\end{eqnarray}
 The integral over $\omega$ can be easily
calculated and we obtain
\begin{eqnarray}
X(i\Omega)=\frac{e^{2}mi}{2\pi ^{2}v_F}\int_{0}^{\infty }
\frac{d\bar{q}}{[1+\bar{q}^{2}]^{3/2}}\nonumber\\
\times\log \left[ 1+\frac{q\Omega}{2me^2v_F\left[1-\frac{1}{\sqrt{1+\bar{q}^{2}}}\right]}\right].
\end{eqnarray}
To obtain the density of states we should substitute $\Omega\rightarrow
-i\epsilon$, where $\epsilon=E-\mu$ is the energy measured from the Fermi
surface, and
take the imaginary part. Thus we obtain 
\begin{equation}
\label{del}
\frac{\delta \nu (\epsilon)}{\nu _{0}}=\frac{e^{2}}{4\pi v_F}\int_{0}^{\infty }
\frac{d\bar{q}}{[1+\bar{q}^{2}]^{3/2}}\log f(\epsilon,\bar{q}),
\end{equation}
where 
\begin{equation}
f(\epsilon,\bar{q})= 1+\frac{x^2\bar{q}^2}{\left[1-\frac{1}{\sqrt{1+\bar{q}^{2}}}\right]^2},
\end{equation}
and
\begin{equation}
x=\frac{\epsilon}{2me^2v_F},
\end{equation}
and $\nu _{0}=m/\pi $ is the DOS
of the noninteracting 2DEG.
For $x\gg 1$ we may put in Eq. (\ref{del})
\begin{equation}
f(x)=2\log x.
\end{equation}
Hence for 
$\left\vert \epsilon\right\vert\gg  (e^2/v_F)E_F$
\begin{equation}
\label{large}
\frac{\delta \nu (\epsilon)}{\nu _{0}}=\frac{e^{2}}{2\pi v_F}
\log\frac{\epsilon}{2me^2v_F}.
\end{equation}

For $x\ll 1$ the main contribution to integral in Eq. (\ref{del}) comes from 
the vicinity of the lower limit, hence we can expand all the functions with respect to $q$, thus obtaining
\begin{equation}
\label{del2}
\frac{\delta \nu (\epsilon)}{\nu _{0}}=\frac{e^{2}}{4\pi v_F}\int_{0}^{\infty }
d\bar{q}\log \left[1+\frac{4x^2}{\bar{q}^2}\right].
\end{equation}
The integral can be easily calculated and we obtain
for $\left\vert \epsilon\right\vert\ll (e^2/v_F)E_F$
\begin{equation}
\label{small}
\frac{\delta \nu (\epsilon)}{\nu _{0}}= \frac{\left\vert \epsilon\right\vert }{8 E_{F}}.
\end{equation}

The equations (\ref{large}) and (\ref{small}) are our main results.  
The correction is smaller
than $\nu _{0}$ consistently with the perturbative assumption. At 
small $|\epsilon|$ our result coincide with that of Ref. \onlinecite{mish} 
up to a
factor $1/2$ (see the statement immediately after Eq. (\ref{x})).  
The linear segment crosses
over to the logarithmic one at the energy scale $|\epsilon|=(e^2/v_F)E_{F}$. 
This last statement  is in good
agreement with numerical results of Ref. \cite{roll}.

We are grateful to E. G. Mishchenko and M. Reizer for their criticism which helped us to find a mistake in our
previous version.

\bibliography{dosm}
\end{document}